\documentclass[times, 10pt,twocolumn]{article} 
\usepackage{latex8}
\usepackage{times}
\newtheorem{theorem}{Theorem}

\newtheorem{definition}{Definition}
\newtheorem{corollary}{Corollary}
\newtheorem{assumption}{Assumption}

\begin{document}

\title{Computing the Equilibria of Bimatrix Games using Dominance Heuristics}

\author{
Raghav Aras \\
LORIA\\
Nancy, France\\
aras@loria.fr
\and Alain Dutech \\
LORIA\\
Nancy, France\\
dutech@loria.fr
\and Fran\c{c}ois Charpillet\\
LORIA\\
Nancy, France\\
charpillet@loria.fr
}
\maketitle
\thispagestyle{empty}
\begin{abstract}
\noindent We propose a formulation of a general-sum bimatrix game as a bipartite directed graph with the objective of establishing a correspondence between the set of the relevant structures of the graph (in particular elementary cycles) and the set of the Nash equilibria of the game.  We show that finding the set of elementary cycles of the graph permits the computation of the set of equilibria.  For games whose graphs have a sparse adjacency matrix, this serves as a good heuristic for computing the set of equilibria.  The heuristic also allows the discarding of sections of the support space that do not yield any equilibrium, thus serving as a useful pre-processing step for algorithms that compute the equilibria through support enumeration.  
\end{abstract}

\section{Introduction}
\noindent Game theory is the study of strategic decision-making.  The  decision-makers are called players.  In bimatrix games, only 2 players are involved.  Each player makes one decision in the game.  This decision pertains to the probability distribution the player conceives over the set of strategies available to him.  As a function of the two decisions, each player receives a real-valued number, called his payoff.  When each player tries to maximize his own payoff, their decisions form an \textit{equilibrium} in which neither player can increase his own payoff by changing his own decision given that the other player sticks to his part of the pair.  Using Brouwer's  fixed-point theorem,  John Nash \cite{nash:equilibrium} proved the existence of such an equilibrium,  since named after him,  for every finite game.  It is equally true that a game may have more than one such equilibrium.
\\The problem of determining the Nash equilibria of a game has occupied much of research in computational game theory (see \cite{mckelveyMcLennan:equilibriaSurvey}, \cite{vonsten:equilibriaSurvey} for excellent surveys).  The principle algorithm for finding a Nash equilibrium of a general-sum game is the Lemke-Howson algorithm (\cite{lemke:lcpAlgorithm}, \cite{lemkeHowson:lcpAlgorithm}).  It solves a linear complementarity program (LCP) formulation of the game.  While in practice this algorithm is quite efficient,  its worst case complexity is exponential \cite{savaniVonStengel:exponentialNash}.  Despite its age,  the algorithm remains the state of the art.  
\\Heuristics about the game structure, therefore, invite interest in finding a sample Nash equilibrium, particularly when the game is large-sized.  It is well-understood that most randomly-generated games allow an equilibrium with \textit{small and balanced} supports \footnote{A support is a subset of strategies that the player uses with positive probability;  balanced here implies that the support size of both players is equal}.  There is theoretical and empirical evidence that in randomly generated bimatrix games (\cite{mclennanBerg:expectedNumberNash}) as $n$, the size of the game increases, the probability of the game having an equilibrium where each player's support has size $n$ becomes vanishingly small.  This heuristic is also used by \cite{dickhautkaplan:supportEnumeration} and \cite{porternudelmanshoham:supportEnumeration} who use a much simpler algorithm than the Lemke-Howson to find a sample Nash equilibrium.  Their algorithm enumerate all support pairs (starting with the smallest-sized ones), and checks if a Nash equilibrium can be formed from a given support pair.  In \cite{porternudelmanshoham:supportEnumeration} it is reporter that in exhaustive computational experience on a variety of games, this algorithm outperforms the Lemke-Howson in finding one Nash equilibrium \cite{porternudelmanshoham:supportEnumeration}.
\\In this paper, we investigate the larger problem of defining a good heuristic for computing the \textit{set} equilibria.  Our approach in general shall be of support enumeration.  Our contribution to this line of research is that we formulate the bimatrix game as a bipartite directed graph that captures the inter-dependencies of the strategies of the game.  We term this graph as the \textit{dominance graph} of the game.  We then establish a correspondence between the \textit{set} of elementary structures of the graph and the set of the equilibria of the game.  In particular,  we show that the set of the elementary cycles of the graph is sufficient to compute the set of equilibria.  Roughly speaking,  we equate a cycle with a support pair.  This heuristic also allows us to \textit{discard} certain portions of support space that will never yield a Nash equilibrium.
\\The motivation for this approach is that graph theory has a predictably large body of work on finding the set of the relevant structures of the graph.  For example,  efficient linear-time algorithms that find the set of elementary cycles (\cite{johnson:elementaryCircuits})or strongly connected components (\cite{tarjan:connectedComponents}, for large-sized, sparsely connected graphs have been known for quite some time.  
\\While support enumeration is a simpler technique to implement than the Lemke-Howson (a given pair of supports can be checked in polynomial time via a linear program if it yields a Nash equilibrium or not), we do state the following caveats.  First,  enumerative methods of LCPs are in general faster than support enumeration (\cite{vonsten:equilibriaSurvey},  \cite{mckelveyMcLennan:equilibriaSurvey}, \cite{audet:EquilibriaEnumeration}) and they require less memory storage.  Second,  a potential drawback of our approach is that if the dominance graph is not sparse, the number of elementary cycles increases faster (with the game size) than support enumeration.  Thirdly,  no efficient algorithm is known that computes only the pair-wise distinct elementary cycles of a graph.  Since two or more cycles composed of the same vertices but in different order could be elementary, there would clearly be a waste if the set of elementary cycles is computed.  
\\The rest of the paper is organized as follows.  In Section \ref{secBimatrixGames} we define bimatrix games, their solutions as well as the support enumeration approach.  Then in Section \ref{secDominanceStructure} we define the formulation the game as a bipartite digraph.  In Section \ref{secEquilibriaCycles} we establish some results about the correspondence between the structures of the graph and the set of equilibria.  Then, in Section \ref{secComputingCycleBasis}, we discuss methods of finding the set of elementary cycles.  We also show how the dominance graph can be feasibly constructed using a more generalized formulation.  Finally in Section \ref{secConclusion}, we summarize this work and discuss its future direction.

\section{Bimatrix Game}
\label{secBimatrixGames}
\noindent In this section we recall standard definitions from game theory.  A bimatrix game $g$ (henceforth, game) is played by two players called player $1$ and player $2$ respectively, and is defined by four elements $g = (M, N, A, B)$.  $M$ and $N$ are the strategy sets of players $1$ and $2$ respectively.  Strategies in $M$ and $N$ are also called \textit{pure} strategies.  Player $1$ has $m$ pure strategies and player $2$ has $n$ pure strategies.  $A$ and $B$ are $m \times n$ matrices and are called respectively player $1$'s and player $2$'s payoff matrix.  If player $1$ chooses strategy $x \in M$  and player $2$ chooses $y \in N$, player $1$ receives the entry $A_{xy}$ as payoff and player $2$ receives the entry $B_{xy}$ as payoff.  
\\\textbf{Notice.}  Henceforth, for convenience, we shall be give definitions and notations only for player $1$ that are also, by obvious analogy, applicable to player $2$, unless we state to the contrary.
\\A \textit{mixed} strategy $p$ for player $1$ is a $m$-column vector where $p_{x}$ represents the probability with which player $1$ chooses the strategy $x \in M$.  In a mixed strategy, pure strategies that receive nonzero probability are said to be in its \textit{support}.  The support of a mixed strategy $p$ shall be denoted by $s_{p}$.  $|s_{p}|$ denotes the size of the set $s_{p}$.  The set of mixed strategies of player $1$ shall be denoted by $\Delta(M)$.  
\\\textbf{Notice.}  Henceforth,  unless specified otherwise:  $x$  shall denote a pure strategy in $M$ and $y$ a pure strategy in $N$,  $A_{x}$ shall denote the $x^{th}$ row vector of $A$ and $B_{y}$ shall denote the $y^{th}$ column vector of $B$.  $p$ shall denote a mixed strategy from the set $\Delta(M)$ and $q$ a mixed strategy from the set $\Delta(N)$.    

\subsection{Dominated Strategies}
\noindent  Strategy $x$ is said to be a \textit{best response} to the strategy $y$ if, $\forall$ $x' \in M\backslash\{x\}$, $A_{xy} \geq A_{x'y}$.  $x$ is said to be a best response to a mixed strategy $q \in \Delta(N)$, if $\forall$ $ x' \in M\backslash\{x\}$, $A_{x} \cdot q$ $\geq$ $A_{x'} \cdot q$.  Given a mixed strategy $q$, the set of pure strategies that are a best response to $q$ is denoted by $BR(q)$.  Finally, the mixed strategy $p$ is said to a best response to the mixed strategy $q$, if $\forall$ $ r \in \Delta(M)\backslash\{p\}$, $p^{T} \cdot A \cdot q$ $\geq$ $r^{T} \cdot A \cdot q$.  Given two strategies, $a$ and $b$, pure or mixed, we shall use $a \rightarrow b$, to mean ``$a$ is a best response to $b$''.  
\\A pure strategy that is not a best response to any pure or mixed strategy is called \textit{strictly dominated}.  Formally, $x$ is strictly dominated, if $\forall$ $q \in \Delta(N)$, $\exists$ $x' \in M\backslash\{x\}$, such that $A_{x'} \cdot q$ $>$ $A_{x} \cdot q$.  The set of strategies of a player can be made smaller by removing from it all strictly dominated strategies.  The following linear program \textbf{LP1} checks if  the pure strategy $x$ is strictly dominated or not.
\begin{itemize}
\item[]  
\parbox[]{3in}{
\textbf{LP1:}\\
\textbf{variables:} $\forall y \in N$, $q_{y}$;  $\epsilon$\\ 
\textbf{maximize:} $\epsilon$\\ 
\textbf{subject to:}\\
\textbf{1.}  $\sum\limits_{y \in N} A_{x'y} q_{y} + \epsilon$ $\leq$ $\sum\limits_{y \in N} A_{xy} q_{y}$, $\forall x' \in M\backslash\{x\}$\\
\textbf{2.}  $\sum\limits_{y \in N}q_{y} = 1$\\
\textbf{3.}  $q_{y} \geq 0$, $\forall y \in N$
}
\end{itemize}
If \textbf{LP1} has a feasible solution and $\epsilon < 0$, then $x$ is strictly dominated.  In other words, if $x$ is not strictly dominated, there exists a mixed strategy $q$ (obtained from the $q$ values in \textbf{LP1}) such that $x \rightarrow q$.   For strictly dominated strategies no such $q$ exists.  Henceforth, we assume that from the sets $M$ and $N$ strictly dominated strategies have been removed by the path-independent process of the iterated elimination of strictly dominated strategies.  

\subsection{Nash Equilibrium}
\noindent We now recall two equivalent definitions of Nash equilibrium \cite{nash:equilibrium},  the central solution concept in game theory.
\begin{definition}
\label{defNashEquilibrium1}
The  mixed strategy pair $(p, q)$ is a Nash equilibrium, if for every mixed strategy $p' \neq p $ of player 1 and every mixed strategy $q' \neq q$ of player $2$,
\begin{eqnarray*}
p^{T} \cdot A  \cdot q  \geq  p'^{T} \cdot A  \cdot q \\
p^{T} \cdot B  \cdot p  \geq  p^{T} \cdot B  \cdot q' 
\end{eqnarray*}
\end{definition}
Thus, if player $2$ is playing $q$, player $1$ cannot improve his payoff by playing a mixed strategy different than $p$.  This is analogously true for player $2$ as well.  The following theorem (\cite{nash:equilibrium}) leads to an equivalent definition of Nash equilibrium.  
\begin{theorem}
\label{theoNash}
The mixed strategy pair $(p, q)$ is a Nash equilibrium, iff $\forall$ $x \in s_{p}$, $x \rightarrow q$ and $\forall$ $y \in s_{q}$, $y \rightarrow p$.
\end{theorem}
Thus, in the Nash equilibrium $(p, q)$, the expected payoff to player $1$ on playing any pure strategy $x \in s_{p}$ when player $2$ chooses his pure strategies according to $q$ is the same, and the expected payoff to player $1$ may be lesser if he uses a pure strategy that lies outside $s_{p}$.  This is analogously true for player $2$ as well.  This implies that strictly dominated strategies cannot be used in the support of any Nash equilibrium.  A Nash equilibrium can be defined in these terms.
\begin{definition}
\label{defNashEquilibrium2}
The mixed strategy pair $(p, q)$ with supports $s_{p}$ and $s_{q}$ respectively is a Nash equilibrium if:
\begin{eqnarray*}
A_{x}^{T} \cdot q = u_{1},  \ \ \forall x \in s_{p}  \\
A_{x'}^{T} \cdot q \leq u_{1}, \ \ \forall x' \in M \backslash  s_{p}\\
p^{T} \cdot B_{y} = u_{2}, \ \ \forall y \in s_{q}\\  
p^{T} \cdot B_{y'} \leq u_{2}, \ \ \forall y' \in N \backslash s_{q}
\end{eqnarray*}
\end{definition}
$u_{1}$ and $u_{2}$ are the expected payoffs for player $1$ and player $2$ respectively for playing the mixed strategies $(p, q)$.

\subsection{Non-Degenerate Games}
\noindent A game is said to be \textit{non-degenerate} if for every mixed strategy,  the number of pure strategies that are a best response to it is less than or equal to the size of its support.  Formally,  
\begin{definition}
\label{defNonDegenerateGame}
Let $g = (M, N, A, B)$ be a bimatrix game.  If for every mixed strategy $p \in \Delta(M)$ and for every mixed strategy $q \in \Delta(N)$,  $|BR(p)|$ $\leq$ $|s_{p}|$ and $|BR(q)|$ $\leq$ $|s_{q}|$, then $g$ is said to be a non-degenerate game.
\end{definition}
A straightforward corollary due to Definition \ref{defNonDegenerateGame} is as follows.
\begin{corollary}
\label{corNonDegenerateGame}
If $(p, q)$ is a Nash equilibrium of a non-degenerate game $g$, then $|s_{p}|$ $=$ $|s_{q}|$ (the supports are said to be balanced).
\end{corollary}
\subsection{Computing Equilibria using Support Enumeration}
\noindent Since the set of mixed strategies of each player is an infinite one,  the set of Nash equilibria may be an infinite one.  For example, in a two-strategy per player game, where the payoff matrices are both the identity matrix, every mixed strategy of one player forms a Nash equilibrium with every mixed strategy of the other player (although the payoffs will be same for every pair).  We are therefore interested in determining a subset $\Omega$ of Nash equilibria that we define as follows.  Let $P(M)$ and $P(N)$ denote the power sets of $M$ and $N$ respectively.
\begin{definition}
\label{defNashEquilibriaSet}
Let $g = (M, N, A, B)$ be a game.  Then, $\Omega$ = $\{s $ : $s$ $\in$ $P(M) \times P(N)$, $\exists$ $(p, q)$ such that $s_{p} \cup s_{q} = s$ and $(p, q)$ is a Nash equilibrium $\}$.
\end{definition}
We are thus interested in determining all the support set pairs that form a Nash equilibrium.  Definition \ref{defNashEquilibrium2} can be directly converted into what \cite{porternudelmanshoham:supportEnumeration} call a \textit{feasibility program}, which is a linear program that accepts as arguments two support sets, $s_{p} \subseteq M$ and $s_{q} \subseteq N$, and checks if they constitute a Nash equilibrium or not.  Since the game involves only two players, the constraints in Definition \ref{defNashEquilibrium2} are all linear.  We denote the linear program corresponding to Definition \ref{defNashEquilibrium2} by \textbf{FP1}.  A simple algorithm to compute $\Omega$ is to run \textbf{FP1} for every pair $(s_{p}, s_{q})$, where $s_{p} \in P(M)$ and $s_{q} \in P(N)$.  There are $(2^{m} - 1)( 2^{n} - 1)$  elements in $P(M) \times P(N)$ and hence this algorithm becomes intractable as $m$ or $n$ grows.
\section{Dominance Graph of the Game}
\label{secDominanceStructure}
\noindent We now develop our idea of deriving a graph from the game $g$ that in such a way that the graph's relevant structural properties (in particular, the number of its elementary cycles) serve as a good heuristic to compute $\Omega$.  We call this graph the game's \textit{dominance graph}.  Each pure strategy is a vertex in this graph.  The graph's adjacency matrix is based on two kinds of sets that we call the \textit{domain} and the \textit{relevancy set} respectively of each pure strategy.  We describe these concepts and the construction of the graph in this section.  The central idea of the construction is that the elements of these two sets represent the vertices of the graph.
\subsection{Domain $D(x)$}
\noindent Theorem \ref{theoNash} says that in a Nash equilibrium mixed strategy pair, each pure strategy of a player is a best response to the other player's mixed strategy.  Therefore a starting point to compute $\Omega$, is to compute for each pure strategy of a player, the set of the mixed strategies to which the pure strategy is a best response.  We define the \textit{domain} of $x$, denoted by $D(x)$, to be the set of subsets of $N$, such that from every element in $D(x)$, a mixed strategy can be formulated to which $x$ is a best response.   
\begin{definition}
The domain of strategy $x$ is the set $D(x) = \{s : $ $s \subseteq N$, $\exists$ $q$ such that $s_{q} = s$ and $x \rightarrow q$\}
\end{definition}
A strategy $x$ that is not strictly dominated may have upto $2^{n} - 1$ elements (subsets of $N$) in its domain.  The domain of a strategy $x$ can be computed by enumerating the elements of the power set of $N$ and checking them individually  to see if they belong to $D(x)$.  We can check via linear programming if, given $s_{1}, s_{0} \subseteq N$,  there exists a mixed strategy $q$ such that $x \rightarrow q$ and $\forall$ $j \in s_{1}$, $q_{j} > 0$ and $\forall$ $k \in s_{0}$, $q_{k} = 0$.  The linear program \textbf{LP2} corresponds to this check.  It takes as arguments the sets $s_{1}, s_{0}$.
\begin{itemize}
\item[]  
\parbox[]{3in}{
\textbf{LP2($s_{1}, s_{0}$):}\\
\textbf{variables:} $\forall y \in N$, $q_{y}$;  $\epsilon$\\ 
\textbf{maximize:} $\epsilon$\\ 
\textbf{subject to:}\\
\textbf{1.}  $\sum\limits_{y \in N} A_{x'y} q_{y} \leq \sum\limits_{y \in N} A_{xy} q_{y}$, $\forall x' \in M\backslash\{x\}$\\
\textbf{2.}  $\sum\limits_{y \in N}q_{y} = 1$\\
\textbf{3.} $q_{j} \geq \epsilon$, $\forall j \in s_{1}$\\
\textbf{4.} $q_{k} = 0$, $\forall k \in s_{0}$
}    
\end{itemize}
Only if \textbf{LP2} has a feasible solution and $\epsilon > 0$, is it true that there exists a mixed strategy $q$ such that $x \rightarrow q$, and $\forall$ $j \in s_{1}$, $q_{j} > 0$ and $\forall$ $k \in s_{0}$, $q_{k} = 0$.  This mixed strategy is obtained from the values of the variables $q_{y}$ (which are the probabilities of pure strategies $y$).  Strictly dominated strategies have empty domains.  
\\$D(x)$ can be computed by enumerating the elements of $P(N)$, the power set of $N$, and executing \textbf{LP2($s_{1}, s_{0}$)} for each element $s \in P(N)$ setting $s_{1} = s$ and $s_{0} = (N - s)$.  The computation of $D(x)$ of each $x \in M$ requires  ($2^{n} - 1$) runs of \textbf{LP2}.  In non-degenerate games, the number \textbf{LP2} runs can be reduced somewhat by keeping track of $(x, q)$ pairs where $s_{q} \in D(x)$.  For example, if $|s_{q}|$ = 2, and we have determined that $s_{q} \in D(x)$, then we do not need to run \textbf{LP2}$(s_{q}, N\backslash s_{q})$ after having found another $x'$ with $s_{q} \in D(x')$.

\subsection{Relevancy Set $R(x)$}
\noindent  The computation of $D(x)$ becomes intractable for large values of $n$.  Besides,  the \textit{size} of $D(x)$ might be too large.  Therefore to use $D(x)$ of each $x$ to obtain the adjacency matrix of the graph is impractical.  In Section \ref{subsecDominanceDomain} we discuss how a subset of $D(x)$ containing supports of small sizes only can be used to construct the graph.  The measure that we shall therefore use to construct the graph is what we term as the \textit{relevancy set} of $x$, denoted by $R(x)$.  It is the set of the pure strategies of player $2$ such that every strategy in $R(x)$ is in the support of \textit{some} mixed strategy to which $x$ is a best response.  
\begin{definition}
\label{defRelevancy}
The relevancy set of strategy $x$ is the set $R(x) = \{y$ : $y \in N$, $\exists$ $q$ such that $q_{y} > 0$ and $x \rightarrow q$\}
\end{definition}
Thus, the relevancy set of $x$ is just the union of the elements of $D(x)$.  The worst-case (as well as best-case) complexity of computing $D(x)$ is exponential.  While computing $R(x)$ has this same worst-case complexity,  its best-case complexity  is much lower.  We describe one method of computing $R(x)$ that works quite well in practice.
\\By the definition of the relevancy set, $R(x)$ is non-empty iff $x$ is non-strictly dominated.  We now show how to determine if a strategy $y^{*} \in N$ is an element of $R(x)$.  In \textbf{LP1}, we make two modifications:  we change constraint \textbf{3.}, to $q_{y} \geq 0$, $\forall y \in N \backslash \{y^{*}\}$, and we add the constraint, $q_{y^{*}}$ = 0.  We call this LP, \textbf{Mod LP1}.  
\\If $x$ is non-strictly dominated then $\epsilon \geq 0$ in \textbf{LP1}.  Therefore, if in \textbf{Mod LP1} (which is the same as \textbf{LP1} but with $q_{y^{*}} = 0$), $\epsilon < 0$, then it implies that $x$ is now strictly dominated only because player $2$ plays $y^{*}$ with zero probability.  In other words, only by re-setting $q_{y^{*}} \geq 0$,  we establish that there exists a mixed strategy $q$ of player $2$ in which he plays $q_{y^{*}}$ with non-zero probability, to which $x$ is a best response.  Thus, we conclude that if $\epsilon < 0$ in \textbf{Mod LP1}, then $y^{*} \in R(x)$.  
\\However, if $\epsilon \geq 0$, it does not rule out that $y^{*}$ could still be in $R(x)$.  By executing \textbf{Mod LP1} $n$ times, each time taking a different strategy from $N$, we might obtain only a subset of $R(x)$.  Assume that this set is $R'(x) \subset N$.  To determine $R(x)$ given $R'(x)$,  we need to execute \textbf{LP2} for either all the pairs $(s_{1}, s_{0})$ (where $s_{1}$ takes on values from the power set of $N - R'(x)$ and $s_{0}$, from the power set of $R'(x)$)  or until $R(x) = N$. \footnote{Other methods too exist for this purpose.  For example,  if $x$ is not strictly dominated, \textbf{LP1} will return a set of pure strategies of player $2$ that are in $R(x)$.  Therefore, only those not found by \textbf{LP1} need to be iteratively checked by \textbf{LP2}.}
\\In view of the worst-case complexity of computing $R(x)$,  for the purpose of the present discussion we make the following simplifying assumption.
\begin{assumption}
\label{assCompletelyConnected}
$\forall$ $x \in M$, $R(x) = N$, and $\forall$ $y \in N$, $R(y) = M$.
\end{assumption}
This assumption does not affect any of the theoretical results we give, but only affects our ability to construct the graph,  an issue we address in Section \ref{subsecDominanceDomain}.  As a matter of fact,  in practice,   we have found that executing \textbf{Mod LP1} on randomly generated bimatrix games,  $n$ times for each $x$,  almost always gives us $R(x) = N$.  Our use $of R(x)$ to describe a graph in the following is purely for expository reasons.  Theorem \ref{theoNash} and  Definition \ref{defRelevancy} give us the following simple corollary.
\begin{corollary}
\label{corRelevancy}
If $(p, q)$ is a Nash equilibrium of a game $g$, then $s_{q} \subseteq R(x)$ for every $x \in s_{p}$ and $s_{p} \subseteq R(y)$ for every $y \in s_{q}$.
\end{corollary}

\subsection{Dominance Graph based on $R(x)$}
\noindent  The relevancy sets $R(x)$ of each $x \in M$, and $R(y)$ of each $y \in N$ as well as the sets $M$ and $N$ lend a certain structure to the game that can be formulated as a \textit{bipartite directed graph} or digraph.  A bipartite digraph is a tuple $G = (U, W, E)$, where $U$, $W$ are finite disjoint sets, and $E$ is a $|U|\times |W|$ matrix called the \textit{adjacency matrix} or the \textit{arc set} such that $E_{uw} = 1$  if an arc exists from $u \in U$ to $w \in W$, and $0$ otherwise.  We say that the arc $(u, w) \in E$, if $E_{uw} = 1$.  The \textit{vertex set} of $G$ is the union of $U$ and $W$.   
\\By setting $U = M$ and $W = N$, and defining $E$ such that $E_{xy} = 1$ iff $y \in R(x)$ and $0$ otherwise, we obtain the dominance graph of the game $g$,  the bipartite digraph $G_{r} = (M, N, E)$.   The vertex set of $G_{r}$ is denoted by $V = M \cup N$.   Thus for every arc in $E$, one endpoint lies in $M$ and the other in $N$ or vice-versa.  $G_{r}$ is a $1$-graph, hence between every ordered pair of vertices, not more than one arc exists.  
\\A bipartite digraph $G = (U, W, E)$ is said to be \textit{completely connected} if for every vertex $u \in U$ and every vertex $w \in W$, the arcs $(u, w)$ and $ (w, u)$ exist in $E$.   Note that if Assumption \ref{assCompletelyConnected} is made,  $G_{r}$ is completely connected.  In a digraph,  the \textit{out-degree} of a vertex $v$, denoted by $O(v)$ is the number of arcs emanating from the vertex, while the \textit{in-degree} of $v$, denoted by $I(v)$  is the number of arcs entering it.  Note that in $G_{r}$, for each $v \in V$,  $O(v) \geq 1$, since the relevancy set of each $v$ is non-empty.  
\subsection{Support Cycle Basis}
\noindent  Some basic structural definitions from graph theory about digraphs that we require to represent the equilibria of the game $g$ in terms of the digraph $G_{r}$ are as follows:  
\\We are given the digraph $G_{r}$ as defined above.  A \textit{path} is a sequence of vertices $(v_{1}\ldots, v_{k})$ such that $\forall$ $1 \leq i < k$, the arc $(v_{i}, v_{i + 1})$ exists in $ E$.  The first vertex $v_{1}$ in the sequence  is called the initial endpoint and  the last vertex $v_{k}$ is called the terminal endpoint.  A \textit{cycle} \footnote{sometimes also called a circuit, esp. in undirected graphs} is a path whose initial endpoint is the same as its terminal endpoint.  An \textit{elementary} (or simple) cycle is a cycle in which no vertex (barring the initial endpoint) occurs twice.  Note that a cycle is a sequence of pure strategies where each strategy is alternatively picked from the two strategy sets.  The \textit{length} of a cycle is the number of vertices in it (not counting the repeating vertex).  A cycle of length $k$ is called a $k$-cycle.  The longest cycle in $G_{r}$ has $2K + 1$ vertices where $K$ is the size of the smaller of the two strategy sets, $M$ and $N$.  We denote the set of elementary cycles of $G_{r}$ by $C_{G_{r}}$.  
\\Henceforth,  for convenience,  we shall refer to elementary cycles as cycles unless we state to the contrary.  The set of vertices that appear in a cycle $\mu$ is denoted by $V(\mu)$ and is called its vertex set.  For $i \in \{1, 2\}$,  the set of vertices (pure strategies) of player $i$ in cycle $\mu$ is denoted by $V_{i}(\mu)$.  Two cycles $\mu$, $\mu'$ are called \textit{equivalent} if $V(\mu) = V(\mu')$, else they are said to be \textit{distinct}.  The set $C_{G_{r}}$ of the elementary cycles of $G_{r}$ can be partitioned into \textit{equivalence} classes, $V^{1}, \ldots, V^{J}$ such that any two cycles from the same class are equivalent and any two cycles from different classes are distinct.  A class is represented by the vertex set of the cycles that belong to that class.  The cycles of a given class are permutations of the vertices of that class.   
\\To eventually be able to compute $\Omega$,  it is enough to know how many equivalence classes there are in $G$ and the definition  (i.e., vertex set) of each class.  It is not necessary to compute the members of each class.  By drawing one member (any member) from each equivalence class, we obtain a set of pairwise distinct cycles.  We define the \textit{support cycle basis} of $G_{r}$ denoted by $\delta_{G_{r}}$ as follows.
\begin{definition}
\label{defCycleBasis}
Let $P(V)$ denote the power set of $V$.  The \textit{support cycle basis} of $G_{r}$ is the set $\delta_{G_{r}} = \{s $ : $s \in P(V)$, $\exists$ $\mu \in C_{G_{r}}$, such that $V(\mu) = s\}$.
\end{definition}
Thus each element of a support cycle basis (henceforth called the cycle basis) \footnote{Not to be confused with the \textit{cycle basis} of a graph which is a set of fundamental cycles of the graph} is a \textit{subset} of $V$.  There exists atleast one cycle whose vertex set equals this subset.  Naturally, it may be possible that other cycles also exist whose vertex set equals this subset.  In the forthcoming discussion we describe a property of cycles that is such that if it is applicable to one cycle, then it is also applicable to every cycle in that cycle's class.  Thus we can refer without ambiguity to an element of $\delta_{G_{r}}$ as a cycle as well a subset.  As we shall see in the next section,  the cycle basis  is important in formulating Nash equilibria in graph-theoretic terms (note the certain similitude between Definitions \ref{defNashEquilibriaSet} and \ref{defCycleBasis}).

\section{Expressing Equilibria as Cycles}
\label{secEquilibriaCycles}
\noindent We now discuss the motivation behind the preceding constructions, that of the dominance graph and the cycle basis.  Our objective in using these two constructs is that they might provide heuristics that enable the computation of the equilibria set $\Omega$ more efficiently than comprehensive support pair enumeration.  We first show how Nash equilibria are related to elementary cycles. 
\\We say that a mixed strategy $(p, q)$ \textit{generates} a given cycle $\mu$ if $V(\mu) \subseteq (s_{p}\cup s_{q})$.  We say that a cycle $\mu$ generates a given mixed strategy $(p, q)$ if $s_{p} = V_{1}(\mu)$ and $s_{q} = V_{2}(\mu)$.
\begin{theorem}
\label{theoNashCycle}
Let $g$ be a game and $G_{r}$ its dominance graph.  Then, for every Nash equilibrium $(p, q)$ of $g$, there exists atleast one cycle $\mu \in \delta_{G_{r}}$ of length $2K + 1$ where $K$ = $\min(|s_{p}|, |s_{q}|)$,  such that $V(\mu)$ $\subseteq$ $(s_{p} \cup s_{q})$.  Moreover, if $|s_{p}|$ $=$ $|s_{q}|$, then $V(\mu)$ $=$ $(s_{p} \cup s_{q})$.
\end{theorem}
\textbf{Proof:}  Assume $(p, q)$ is a Nash equilibrium of $g$.  Denote $s_{p}$ by $s_{1}$ and $s_{q}$ by $s_{2}$.  $\forall$ $i \in \{1, 2\}$,  let $L_{i}$ be a stack in which the elements of $s_{i}$ have been pushed in any order.  Let $K$ denote the size of the smaller of the two stacks and $k$ the subscript of that stack.  Let $\mu$ be a list.  The $j^{th}$ element of $\mu$ is denoted by $\mu_{j}$.  Now, remove $K$ elements from $L_{k}$ and $L_{-k}$ \footnote{$-k$ denotes ``not $k$''} each, and place them \textit{alternatively} in $\mu$ with an element of $L_{k}$ being $\mu_{1}$.  By Corollary \ref{corRelevancy}, $\forall$ $i \in \{1, 2\}$, $\forall$ $u \in L_{i}$, $s_{-i} \subseteq R(u)$ and thus, $\forall$ $v \in s_{-i}$, $(u, v) \in E$.  Therefore,  for every $1 \leq j < 2K$,  the arc $(\mu_{j}, \mu_{j + 1})$ is an element of $E$.  Since, $\mu_{1} \in L_{k}$ and there are $2K$ elements in $\mu$, $\mu_{2K} \in L_{-k}$.  But the arc $(\mu_{2K}, \mu_{1})$ exists in $E$ (by Corollary \ref{corRelevancy}).  Therefore, adding the element $\mu_{1}$ at position $2K + 1$, gives us an elementary cycle,  as claimed in the first statement.  
\\The vertex set of $\mu$ is $L_{k} \cup L_{-k}(K)$, where $L_{-k}(K) \subseteq L_{-k}$ such that $|L_{-k}(K)|$ = $K$.  If (as in a non-degenerate game) $|L_{k}|$ $=$ $|L_{-k}|$ $=$ $K$,  the elementary cycle $\mu$, as constructed above, has a vertex set that equals $L_{k} \cup L_{-k}(K)$ $=$ $L_{k} \cup L_{-k}$ as claimed in the second statement. $QED$.
\\Since the order in which the vertices are put in the stacks does not matter in the proof of Theorem \ref{theoNashCycle},  it follows that $(p, q)$ generates every cycle of the class to which $\mu$ belongs.  This allows us, as stated before, to refer to an element of $\delta_{G_{r}}$ as a cycle as well as a subset.  We can also refer to the cycle generated by $(p, q)$.

\subsection{Support Trees}
\noindent  Theorem \ref{theoNashCycle} implies that every Nash equilibrium of a game generates a cycle.  If the game is non-degenerate or if the supports of the equilibrium are balanced, then the cycle also generates the Nash equilibrium.  In particular,  \textit{every $3$-cycle of $G_{r}$ generates a Nash equilibrium of $g$}.  We cannot generalize this statement, however.  That is, not every Nash equilibrium of $g$ can be generated by a cycle of size $\geq$ $5$.
\begin{corollary}
\label{corCycleNash}
Given a game $g$ and its dominance graph $G_{r}$, and a Nash equilibrium $(p, q)$ of $g$ that generates the cycle $\mu \in \delta_{G_{r}}$ of length $\geq$ $5$.  Then  it is possible that $\mu$ does not generate $(p, q)$, that is, it is possible that $s_{p} \neq V_{1}(\mu)$ or $s_{q} \neq V_{2}(\mu)$.
\end{corollary}
Due to this corollary,  it would appear that computing the cycle basis may not be sufficient to compute the set of Nash equilibria $\Omega$.  However,  as we describe in the following,  while a cycle itself may not generate a particular equilibrium,  a cycle and an auxiliary set of $3$-cycles would generate that equilibrium.  Note that every $3$-cycle is necessarily elementary.  
\\Consider a mixed strategy $(p, q)$ that is a Nash equilibrium.  By Theorem \ref{theoNashCycle} it generates a cycle.  Let this cycle be $\mu \in \delta_{G_{r}}$.  If $|s_{p}|$ $=$ $|s_{q}| = K$, then the cycle generates the equilibrium as well.  So, the case that requires generalization is if $|s_{p}|$ $\neq$ $|s_{q}|$.  Let $|s_{p}|$ $<$ $|s_{q}|$, and let $s = s_{q} - V_{2}(\mu)$ ($s$ contains the pure strategies in $s_{q}$ that have not been ``used up'' in $\mu$).  By Corollary \ref{corRelevancy}, for each $v \in s$, and for each $u \in V_{1}(\mu)$, the arcs $(u, v)$ and $(v, u)$ exist in $E$.  Thus, each element $v \in s$ forms the cycle $(v, u, v)$ with atleast one vertex of $u \in V_{1}(\mu)$.  
\\Therefore,  every Nash equilibrium $(p, q)$ is such that the union of its support sets equals the union of the vertex sets of a set of cycles $\tau(p, q)$ where each cycle is from $\delta_{G_{r}}$.  In this set,  there is a cycle $\mu$ of length $2K + 1$ and some other $3$-cycles, whose vertex sets have one element in common with the vertex set of $\mu$.  We call this set of cycles a \textit{support tree} (henceforth, tree) \footnote{A structure such as $\tau(p, q)$ is a \textit{tree} of the underlying undirected graph of $G_{r}$}.  We say that the mixed strategy $(p, q)$ generates the tree $\tau(p, q)$ if the latter is obtained is the manner just described.  Therefore if $(p, q)$ generates $\tau(p, q)$, $(s_{a} \cup s_{q})$ = $\bigcup\limits_{c \in \tau(p, q)}V(c)$.  Moreover, as in the case of cycles and balanced supports,  here $\tau(p, q)$  generates $(p, q)$ as well.  We can thus find a tree of $G_{r}$ that generates a given Nash equilibrium (In the case of an equilibrium with balanced supports, there are no $3$-cycles in $\tau$).
\begin{theorem}
\label{theoNashTree}
Let $g$ be a game and $G_{r}$ its dominance graph.  Every Nash equilibrium $(p, q)$ of $g$ generates atleast one support tree $\tau(p, q)$ of $G_{r}$.
\end{theorem}   
An important consequence of the two preceding theorems is that, we can use them for deciding if a certain strategy is eliminable i.e., it does not occur in the support of any Nash equilibrium.  More generally, we can use the theorems to discard a subset of strategies, if we find that they do not yield any cycle.

\section{Computing the Support Cycle Basis}
\label{secComputingCycleBasis}
\noindent The two theorems of the last section establish that the set of Nash equilibria $\Omega$ can be computed from the cycle basis only.  So a general scheme to compute $\Omega$ that we call \textit{support tree enumeration} as follows.  We first determine the cycle basis $\delta_{G_{r}}$ from $C_{G_{r}}$ which also gives us all $3$-cycles.  Denote the set of $3$-cycles by $\delta_{G_{r}}^{3}$, and by $P^{3}$ its power set.  The set of support trees is obtained by keeping those elements of $(\delta_{G_{r}} - \delta_{G_{r}}^{3}) \times P^{3}$ that satisfy the definition of a support tree.  Finally, for each cycle or support tree $\mu$ found, we run \textbf{FP1} with arguments $V_{1}(\mu)$ and $V_{2}(\mu)$.  
\\The cycle basis is just a set of cycles of $G_{r}$.  The problem of determining the set of elementary cycles of a directed graph is a well studied one in graph theory.  To our knowledge,  the algorithm due to Johnson \cite{johnson:elementaryCircuits} is the most efficient in this regard.  Its run-time is bounded by $O((v + e)(c + 1))$, where $v$ is the number of vertices,   $e$ the number of arcs and $c$ the number of elementary cycles of the graph.  It computes the set $C_{G_{r}}$.  We do not know of any algorithm that computes efficiently the subset $\delta_{G_{r}}$ of $C_{G_{r}}$.  
\\Johnson's algorithm detects the strongly connected components (SCCs) of a digraph $G$ and then finds all the elementary cycle of each SCC.  An SCC is a subset $V'$ of the vertices of $G$ such that for every pair of vertices $u, v, \in V'$ there exists an elementary path of vertices of $V'$ such that its initial endpoint is $u$ and terminal endpoint is $v$.  There exist efficient, linear-time algorithms that find all the SCCs of digraph.  The efficiency of Johnson's algorithm depends on the density of the matrix $E$ (and on the number of SCCs;  the more SCCs, the better it is).
\\A completely connected bipartite graph  has just one SCC and has the maximum number of elementary cycles that a graph of its size (in the number of vertices, say $k$) can have.  This number (\cite{johnson:elementaryCircuits}), grows, faster than the number of total supports of the game as $k$ grows.  Therefore,  constructing $G_{r}$ by making Assumption \ref{assCompletelyConnected} and then enumerating its cycles using Johnson's algorithm (or any other) is guaranteed to be worse than enumerating the elements of $P(M) \times P(N)$.  On the other hand, as noted before, $R(x)$ is difficult to compute as well.  
\\So, we would like to construct a graph without the (forced) complete connectedness of $G_{r}$ but without actually computing $R_{x}$.  Moreover,  we would like Theorems \ref{theoNashCycle} and \ref{theoNashTree} to be true for this graph as well.  We now describe how a graph that is based on $D(x)$ can satisfy these criteria.

\subsection{Dominance Graph based on $D(x)$}
\label{subsecDominanceDomain}
\noindent  We define a dominance graph based on the domain, denoted by $G_{d}$ as the bipartite digraph $G_{d} = (P(M), P(N), E)$.  In this graph a vertex is an element of the power set of the set of pure strategies and thus corresponds to either a pure strategy or to a set of pure strategies.  In $G_{d}$  an arc is made from a vertex $s$ to the vertex $t$, if every pure strategy in $s$ is a best response to some mixed strategy with support $t$.  Thus, given $s \in P(M)$ and $t \in P(N)$, $E_{st} = 1$, iff, $\forall$ $x \in s$, $t \in D(x)$. Using Theorem \ref{theoNash},  it can be verified that all $3$-cycles of $G_{d}$ are Nash equilibria, just as the $3$-cycles of $G_{r}$ are.  Additionally, in $G_{d}$ only $3$-cycles generate Nash equilibria.  Cycles of longer lengths need not be considered.
\\It is easy to see the motivation behind the construction of $G_{d}$.  Even if the sets $R(x)$ are considered as given, $G_{r}$ contains a lot of superfluous information in terms of the arcs it contains.  For example,  let $x$ be a vertex in $G_{r}$, and let the arcs $(x, y)$ and $(x, w)$ exist in $E$ of $G_{r}$.  Suppose that $x$ is not a best response to $y$,  $x$ is not a best response to $w$ but $(y, w) \in D(x)$.  This implies, that $x$ is a best response to a mixed strategy $q$ that has in its support $y$,  if and only if $w$ (or some other pure strategies) also occur in the support.  A graph such as $G_{d}$ contains more precise information.       
\\Thus on the one hand, we have the small but quasi-completely connected graph $G_{r}$, and on the other, the very large, but possibly sparsely connected graph $G_{d}$.  We can therefore seek to construct a dominance graph that is an \textit{intermediate} between $G_{r}$ and $G_{d}$.  The intermediacy is in the size of the two vertex sets of $G_{i}$.  In $G_{d}$, they are $P(M)$ and $P(N)$.  In general, they can be any subsets of $P(M)$ and $P(N)$.  We define an intermediate graph denoted by $G_{i}$ as $G_{i} = (P_{k}(M), P_{l}(N) , E)$ where $P_{k}(M) \in P(M)$ consists only of elements of $P(M)$ of size $k$ or less and $P_{l}(N) \in P(N)$ consists only of elements of $P(N)$ of size $l$ or less.  The definition of $E$ requires some care.  Given an $s \in P_{k}(M)$ we define all the outgoing arcs from $s$ as follows. (by analogy, the following discussion is also applicable for every $t \in P_{l}(N)$ ).  Let $D_{l}(x) \subseteq D(x)$ denote the subset of the domain of $x$ such that its elements are of size $l$ or less.  For example, the $D_{2}(x)$ can contain only pure strategies and pairs of pure strategies from $N$.  Let $S(v)$ denote the set of pure strategies in the vertex $v$ and $L(v)$ the size of $S(v)$.   Then,    
\begin{enumerate}    
\item Let $T(s) = \{t \in P_{l}(N)$ : $\forall$ $x \in s$, $t \in D_{l}(x)\}$.  Then, $\forall$ $t \in T(s)$, $E_{st} = 1$
\item If $T(s)$ is empty,  let $D(s)$ = $\{t \in P_{l}(N)$ : $S(t)$ = $\bigcap\limits_{x \in s} D_{l}(x)$$\}$. Then, $\forall$ $t$ such that $t = \arg\max\limits_{L(w)}\{w \in D(s)\}$, $E_{st} = 1$.
\item If $T(s)$ and $D(s)$ are both empty, then $\forall$ $t \in (\bigcup\limits_{x  \in s}D_{l}(x))$, $E_{st} = 1$.
\item If $s = x$ (i.e., it is a pure strategy) and $D_{l}(x)$ is empty, then $\forall$ $t \in P_{l}(N)$, $E_{st} = 1$.
\end{enumerate} 
In $G_{d}$ only Case 1 is needed.  In Cases $2$, $3$ and $4$, we are creating \textit{artificial arcs}.  These are needed, since in $G_{i}$,  we disallow vertices $v \in P_{l}(N)$ such that $L(v) > l$.  Thus, it is possible that for a given $s$, there is no outgoing arc (the vertex is isolated) using just Case 1.  This would happen either because $D_{l}(x)$ is empty for some $x \in s$ or that none of $x$ in $s$ have a common element in their domains $D_{l}$.  Hence we need Cases $2$, $3$ and $4$.  If the values of $l$ is small,  the computation of $D_{l}(x)$ is tractable.  The intuition behind the definition of $E$ above, is that in most games  even for small values of $l$,  Cases $2$, $3$ and $4$ are not needed, and hence artificial arcs (that introduce artificial cycles into $G_{i}$) need not be made.  We define a support tree $\tau$ of $G_{i}$ to be a set of cycles such that each pair of cycles in it has one vertex in common.  The definition of $G_{i}$ leads to the following theorem.
\begin{theorem}
\label{theoNashCycleInter}
Given an intermediate dominance graph $G_{i} = (P_{k}(M), P_{l}(N), E)$ of a game $g$, every Nash equilibrium $(p, q)$ of $g$ generates atleast one support tree $\tau$ of $G_{i}$ and $\tau$ generates  $(p, q)$ as well.
\end{theorem}
We summarize a general scheme to compute the set $\Omega$ as follows.  For players $1$ and $2$, we set the values $k$ and $l$ respectively, to fix the sizes of $P_{k}(M)$ and $P_{l}(N)$.  Then, using \textbf{LP2}, we compute $\forall$ $x \in M$,  $\forall$ $1 \leq i \leq k$, $D_{i}(x)$, and $\forall$ $y \in N$,  $\forall$ $1 \leq i \leq l$, $D_{i}(y)$.  We then fill the entries of the matrix $E$ as described in Section \ref{subsecDominanceDomain} to obtain the graph $G_{i}$.  We then use an elementary cycle-finding algorithm such as \cite{johnson:elementaryCircuits} to get the cycle basis of $G_{i}$.  We then find sets of cycles from $\delta_{G_{i}}$ that contain pairwise intersecting cycles.  Then the program \textbf{FP1} is run for each such set to obtain $\Omega$.

\subsection{Results}
\noindent  Table \ref{tabCycleBasis} shows some preliminary results (the games were generated by the GAMUT software \cite{porternudelmanshoham:gamut})
about the sizes of the cycle basis in $G_{r}$.  The relevancy sets were obtained through \textbf{Mod LP2}, and in fact in all cases, the relevancy sets equaled the other player's strategy set.  We did not use Johnson's algorithm since it does not directly compute the cycle basis.  For our purpose, we have conceived a simple enumerating algorithm that builds elementary cycles of length $k + 1$ from those of length $k$.  The different cycle lengths to be considered are $k = 3, 4, \ldots (2K + 1)$.  Each elementary $k$-cycle of length $k$ is stored in a vector.  Before storing, the vertices are sorted, and converted into a number using a coding scheme.  In the first step,  all $3$-cycles are computed by a simple search.  When searching for a $(k + 1)$-cycle from a $k$-cycle, an expansion is done (using fixed look-ahead) only if the resulting cycle is not already present in the set of $(k + 1)$-cycles.  
\begin{table}
\begin{center}
{\caption{Average size of $\delta_{G_{r}}$ in random bimatrix games of sizes $7$ to $11$.  $S = P(M) \times P(N)$}
\begin{tabular}{|l|c|c|c|c|c|}
\hline
 $m = n =$   & $7$ & $8$ & $9$ & $10$ & $11$\\ 
\hline
$|\delta_{G_{r}}|$     & 757 &  3775 & 11772 & 48768 & 252567\\ 
\hline 
$|\delta_{G_{r}}| / |S|$     & 0.04 & 0.058  & 0.045 & 0.046 & 0.06\\ 
\hline 
$T (secs)$     & $< 1$ & $< 1$   & 5.3  & 9.8  & 67 \\ 
\hline 
\end{tabular}
\label{tabCycleBasis}}
\end{center}
\end{table}    
We also conducted several experiments to generate statistics about the domain set for a variety of $10 \times 10$ games using GAMUT.  We do not report our findings here for want of space, but we do mention that (predictably) for random games, the adjacency matrix of $G_{d}$ is very dense (about 65$\%$ of entries are $1$).  Games with sparse matrices were ``WarOfAttrition''(15$\%$), ``LocationGame''(10$\%$) and ``GuessTwoThirdsAve''(15$\%$) among others.  
\section{Conclusion and Future Work}
\label{secConclusion}
\noindent We have presented a heuristic for the computation of the set of equilibria of bimatrix games as well as for identifying eliminable strategies (those that are not in any Nash equilibrium).  We have formulated the heuristic in graph-theoretic terms with the idea that certain games can be converted to sparsely connected digraphs, which can then be mined for interesting structures.  In this paper, we showed that we can re-design a game to be a digraph whose elementary cycles can be checked directly to see if they yield Nash equilibria.  The bulk of the paper concerned graphs conceived with the relevancy set.  As we stated, $G_{r}$ was used mainly for expository purposes.  Our immediate work concerns more focused computational experience with intermediate graphs $G_{i}$.  At the present time,  there are not many approaches in the literature for computing the set of Nash equilibria, and we hope that our approach is a useful contribution.
\\\textbf{Acknowledgements.}  We are thankful to Martin Allen and Vishesh Vikas for helpful discussions.
\nocite{*}
\bibliographystyle{latex8}
\bibliography{CameraReadyICTAI06}

\end{document}